\documentclass[12pt]{article}
\usepackage{graphicx}

\begin{document}

\baselineskip=18pt plus 1pt minus 1pt

\begin{center}

{\large \bf
E(5), X(5), and prolate to oblate shape phase transitions in relativistic 
Hartree Bogoliubov theory} 

\bigskip\bigskip

{R. Fossion$^{a,b}$\footnote{e-mail: ruben.fossion@pd.infn.it},
Dennis Bonatsos$^{c}$\footnote{e-mail: bonat@inp.demokritos.gr},
and G. A. Lalazissis$^{a}$\footnote{e-mail: glalazis@auth.gr}
}

\bigskip

{$^{a}$ Physics Department, Aristotle University of Thessaloniki}

{GR-54124 Thessaloniki, Greece} 

{$^{b}$ Istituto Nazionale di Fisica Nucleare and 
Dipartimento di Fisica Galileo Galilei}

{Via Marzolo 8, I-35131 Padova, Italy} 

{$^{c}$ Institute of Nuclear Physics, National Center for Scientific Research 
``Demokritos''}

{GR-15310 Aghia Paraskevi, Attiki, Greece}

\end{center}

\bigskip\bigskip
\centerline{\bf Abstract} \medskip 
Relativistic mean field theory with the NL3 force is used for producing 
potential energy surfaces (PES) for series of isotopes suggested as exhibiting 
critical point symmetries. Relatively flat PES are obtained for nuclei 
showing the E(5) symmetry, while in nuclei corresponding to the X(5) 
case, PES with a bump are obtained. The PES corresponding to the Pt chain of 
isotopes suggest a transition from prolate to oblate shapes at $^{186}$Pt.   

\bigskip\bigskip 

\section{Introduction} 

Critical point symmetries \cite{IacE5,IacX5} corresponding to shape phase 
transitions in nuclear structure are recently receiving considerable attention,
since they lead to parameter independent (up to overall scale factors) 
predictions for spectra and $B(E2)$ transition rates which compare well with 
experimental data \cite{CZE5,CZX5,ClarkE5,ClarkX5}. 
The E(5) \cite{IacE5} and X(5) \cite{IacX5} critical point symmetries
correspond to special solutions of the Bohr Hamiltonian \cite{Bohr}, in both 
of which an infinite square well potential in the quadrupole ($\beta$) 
degree of freedom is assumed. In E(5), corresponding to the 
transition from vibrational [U(5)] to $\gamma$-unstable [O(6)] nuclei, 
a $\gamma$-independent potential is assumed \cite{Wilets}, while in X(5),
related to the transition from vibrational [U(5)] to axially symmetric 
prolate [SU(3)] nuclei, a potential of the form $u(\beta)+u(\gamma)$ 
is used. A systematic study of phase transitions in nuclear collective models 
has been given in \cite{RoweI,RoweII,RoweIII}. 
 
It is of interest to examine if the assumption of a relatively flat potential
in the E(5) and X(5) models is justified through the use of a completely 
different method, as the Relativistic Mean Field (RMF) approach \cite{VALR.05}
using the NL3 effective force, the parameter set of which has been fixed 
by fitting ground state properties of spherical nuclei \cite{LKR.97}. 
The calculation of Potential Energy Surfaces (PES) for a series 
of isotopes in which a critical nucleus appears should result in a relatively 
flat PES for this particular nucleus. 

Furthermore, the question of a prolate to oblate shape phase transition, 
expected to take place in the Pt region, has been raised recently 
\cite{Linnemann}. RMF theory \cite{VALR.05} is a suitable tool for locating 
the prolate to oblate shape transition in the Pt series of isotopes. 

The Relativistic Hartree Bogoliubov model used in the present work is briefly 
described in Section 2. In this variation of the RMF theory, pairing 
correlations are treated in a self consistent manner, since in constrained 
calculations the correct treatment of pairing is essential for a reliable 
description of the potential landscapes. RMF theory is then used to produce 
potential energy surfaces for chains of isotopes 
involving nuclei suggested to be good examples of the E(5) and X(5) critical 
point symmetries in Sections 3 and 4 respectively, while in Section 5 a study
of isotopic chains related to the prolate to oblate shape transition is given. 
The main conclusions arising from the present results and plans for further 
work are finally discussed in Section 6.
 
\section{The Relativistic Hartree Bogoliubov model} 

The model describes the nucleus as a system of Dirac nucleons which interact
in a relativistic covariant manner through the exchange of virtual mesons.
The relativistic extension of the Hartree Fock Bogoliubov theory is described
in detail in Refs. \cite{Kucha.91,LA.99}.
The generalized single-particle Hamiltonian of HFB theory contains two average
potentials: the self consistent field $\hat{\Gamma}$ which encloses all the 
long range {\em ph} correlations, and the pairing field $ \hat{\Delta}$ which 
sums up the {\em pp}-correlations. In the Hartree approximation for the 
self-consistent mean field, the Relativistic Hartree-Bogoliubov equations read
\begin{eqnarray}
\label{equ.2.2}
\left( \matrix{ \hat h_D -m- \lambda & \hat\Delta \cr
        -\hat\Delta^* & -\hat h_D + m +\lambda} \right)
\left( \matrix{ U_k({\bf r}) \cr V_k({\bf r}) } \right) =
E_k\left( \matrix{ U_k({\bf r}) \cr V_k({\bf r}) } \right), 
\end{eqnarray}
where $m$ is the nucleon mass and  $\hat h_D$ is the single-nucleon Dirac
Hamiltonian 
\begin{equation}
\label{statDirac}
\left\{-i\mbox{\boldmath $\alpha$}
\cdot\mbox{\boldmath $\nabla$}
+\beta(m+g_\sigma \sigma)
+g_\omega \omega^0+g_\rho\tau_3\rho^0_3
+e\frac{(1-\tau_3)}{2} A^0\right\}\psi_i=
\varepsilon_i\psi_i, 
\end{equation}
with $\sigma$, $\omega$, and $\rho$ the meson fields,  A  the
electromagnetic potential and $g_\sigma$, $g_\omega$, and $g_\rho$ the 
corresponding coupling constants of the mesons to the nucleon.  

In Eq. (\ref{equ.2.2})  the chemical potential $\lambda$  has to be determined 
by the particle number subsidiary condition in order that the
expectation value of the particle number operator
in the ground state equals the number of nucleons.
$\hat\Delta $ is the pairing field. The column
vectors denote the quasi-particle spinors and $E_k$
are the quasi-particle energies.
The self-consistent solution of the Dirac-Hartree-Bogoliubov
integro-differential eigenvalue equations and Klein-Gordon equations for the
meson fields determines the nuclear ground state.
In the present variation of the model these equations are solved by 
expanding the nucleon spinors $U_k({\bf r})$ and $V_k({\bf r})$,
and the meson fields, in terms of the eigenfunctions
of a deformed axially symmetric oscillator potential.
The calculations have been performed using the NL3 Lagrangian parametrization
\cite{LKR.97}, which has been proved very successfull in describing various 
nuclear properties at and away from the line of $\beta$-stability. Finally, 
for the pairing field
we employ the pairing part of the Gogny interaction \cite{BGG.84}.
The basic advantage of this choice is the fact that the Gogny force has a 
finite range which automatically guarantees a proper cut-off in momentum space.

\section{The E(5) shape phase transition} 

The first nucleus to be identified as exhibiting E(5) behaviour was 
$^{134}$Ba \cite{CZE5}, while $^{102}$Pd \cite{Zamfir,Kalyva} also seems 
to provide a very good candidate. Further studies on $^{134}$Ba \cite{AriasE2}
reinforced this conclusion, while $^{108}$Pd has been suggested \cite{Zhang}
to be an E(5) candidate and 
$^{100}$Pd was found \cite{E5} to 
correspond to a numerical solution using a $\beta^4$ potential 
\cite{Ariasb4,Ariasb4b} instead of an infinite square well potential in 
the E(5) framework. A systematic search
\cite{ClarkE5,Kirson} on available data on energy levels and B(E2) transition 
rates also singled out $^{128}$Xe as a good candidate.
$^{130}$Xe has also been suggested \cite{Liu} as a possible candidate, in 
agreement with a recent report \cite{Kneissl} on measurements of E1 and M1 
strengths of $^{124-136}$Xe, which provide evidence for a shape phase 
transition around $A\simeq 130$. 

Potential energy surfaces for $^{96-114}$Pd, $^{118-134}$Xe, and 
$^{118-138}$Ba are shown in Figs. 1-3.  
$^{100}$Pd and $^{102}$Pd show PES which are quite flat. The PES of $^{108}$Pd 
and $^{110}$Pd are also quite flat. 
In addition, it is clear that $^{126-130}$Xe do exhibit flat PES, being on the
way from vibrational behaviour, seen in $^{134}$Xe ($R_4=2.044$ \cite{NDS}), 
to $\gamma$-unstable behaviour in the lighter Xe isotopes \cite{IBM}. 
Similarly, $^{132-134}$Ba exhibit rather flat PES, being on the way from 
vibrational behaviour, seen in $^{136}$Ba ($R_4=2.280$ \cite{NDS}), 
to $\gamma$-unstable behaviour in the lighter Ba isotopes.

In general, we remark that the assumption of a rather flat potential
in the E(5) critical point symmetry is supported quite well by the present 
calculations in the nuclei which have been suggested as good E(5) examples. 

\section{The X(5) shape phase transition} 

The first nucleus to be identified as exhibiting X(5) behaviour was 
$^{152}$Sm \cite{CZX5}, followed by $^{150}$Nd \cite{Kruecken}. 
Further work on $^{152}$Sm \cite{ZamfirSm,Clark,CZK,Bijker} and $^{150}$Nd
\cite{Clark,CZK,Zhao} reinforced this conclusion. The neibouring N=90 
isotones $^{154}$Gd \cite{Tonev,Dewald} and $^{156}$Dy \cite{Dewald,CaprioDy}
were also seen to provide good X(5) examples, the latter being of less
good quality. A systematic study \cite{ClarkX5} 
of available experimental data on energy levels and B(E2) transition rates 
suggested $^{126}$Ba as a possible good candidate, in 
addition to the N=90 isotones of Nd, Sm, Gd, and Dy. 
$^{122}$Ba has also been recently suggested \cite{Fransen} as a possible
candidate.  In parallel, $^{148}$Nd has been suggested as corresponding to
an analytic solution \cite{X5}  using a  $\beta^2$ potential instead of an 
infinite square well potential within the X(5) framework. 
 
Potential energy surfaces for $^{144-156}$Nd, $^{146-158}$Sm, $^{148-156}$Gd,
and $^{150-158}$Dy are shown in Figs. 4-7, while PES for $^{118-138}$Ba 
have already been exhibited in Fig. 3. We remark that for all of the 
above mentioned nuclei [$^{122,126}$Ba, $^{150}$Nd, $^{152}$Sm, $^{154}$Gd, 
$^{156}$Dy] the PES is not flat, exhibiting a deeper minimum 
in the prolate ($\beta_2>0$) regime and a shallower minimum in the
oblate ($\beta_2<0$) region. Relatively flat PES occur for the N=86
nuclei $^{146}$Nd, $^{148}$Sm, $^{150}$Gd, and $^{152}$Dy (to a lesser 
extend). 

The present results for $\beta_2>0$ are in good agreement with the PES obtained
for $^{144-158}$Sm by Relativistic Mean Field (RMF) theory in Ref. \cite{Meng},
using the NL3 force, as well as the NL1, NLSH and TM1 forces. In Ref. 
\cite{Meng} the PES for $^{152}$Sm, as well as for neighbouring nuclei, 
presents a single minimum in the region $\beta_2>0$, in agreement with Fig. 5~.
Furthermore the present results for $^{146-156}$Sm and $^{154}$Gd are in good 
agreement with Nilsson-Strutinsky-BCS calculations \cite{ZhangSm}. 

From the above we conclude that the present results, in agreement with earlier 
calculations, do not predict flat PES for the N=90 isotones, which are 
the best experimental manifestations of X(5). However, 
the existence of a bump in the PES corresponding to good experimental 
examples of X(5) might be related to the 
success of the confined $\beta$-soft (CBS) rotor model \cite{Pietr1,Pietr2},
employing an infinite square well potential displaced from zero, as well as 
to the relevance of Davidson potentials \cite{Dav,Rowe} of the form 
$\beta^2 + \beta_0^4/\beta^2$ (where $\beta_0$ is the minimum of the potential)
in the description of X(5) properties \cite{varPLB,varPRC}.
It can also be related to the significant five-dimensional centrifugal effect
\cite{Caprio72}, found recently through the use of novel techniques 
allowing for the exact numerical diagonalization of the Bohr Hamiltonian 
\cite{Rowe735,Rowe45,Rowe753}.

\section{The prolate to oblate transition} 

The chain of nuclei $^{180}$Hf, $^{182,184,186}$W, $^{188,190,192}$Os, 
$^{194,196}$Pt, and $^{198,200}$Hg has been suggested \cite{Linnemann}, 
on the basis of experimental data, as exhibiting a transition from prolate 
to oblate shapes at $^{194}$Pt. The PES shown in Fig. 8 clearly show such 
a transition, since the prolate minimum is lower than the oblate minimum 
up to $^{192}$Os, while in $^{194}$Pt the opposite picture appears. 
The PES for $^{194}$Pt and especially for $^{196}$Pt is flat, evolving 
in $^{198}$Hg and $^{200}$Hg towards a vibrational shape. 
The quadrupole moments of the nuclei belonging to this chain, shown 
in Fig. 9, also exhibit the transition from negative values (corresponding 
to prolate shapes) to positive values (corresponding to oblate shapes). 
It should be noticed that in Fig. 9 (as well as in similar subsequent 
figures) the experimental values should be compared to the theoretical 
results for the total quadrupole moments, while the theoretical results 
for the separate contibutions of the protons and of the neutrons are shown 
for completeness. It is, however, worth remarking that the experimental values
are close to the theoretical contributions of the protons.   

It should be noticed, however, that the PES of the $^{184-202}$Pt isotopes, 
shown in Fig. 10, exhibit a transition from prolate to oblate shapes 
between $^{186}$Pt (prolate) and $^{188}$Pt (oblate). In $^{184-192}$Pt
two minima appear, the prolate minimum being lower than the oblate minimum 
in $^{184,186}$Pt, with the opposite situation occuring in $^{188-192}$Pt. 
Beyond $^{194}$Pt the PES becomes flat, evolving towards a vibrational 
shape, reached at $^{202}$Pt. The quadrupole moments of $^{184-202}$Pt, 
shown in Fig. 11, also exhibit a transition from prolate to oblate 
behaviour between $^{186}$Pt and $^{188}$Pt.  

The prediction that the nucleus $^{186}$Pt is critical is supported by 
several pieces of evidence. As seen in Fig. 12, the $\beta_1$-bandheads 
(normalized to the energy of the $2_1^+$ state) exhibit a minimum 
at $^{186}$Pt, while the crossover of the (normalized to the energy of 
the $2_1^+$ state) bandheads of the $\beta_1$ and $\gamma_1$ bands 
also occurs at the same nucleus. Furthermore, mapping the Pt isotopic chain 
on the symmetry triangle \cite{Casten} of the Interacting Boson Model 
\cite{IBM} shows \cite{McCPt} that $^{186}$Pt lies very close to the shape 
phase coexistence region of IBM \cite{IZC,McCZC}. 

A similar transition from prolate to oblate shapes is seen in the 
$^{188-200}$Os chain, shown in Fig. 13. The prolate minimum is lower than 
the oblate minimum in $^{188-192}$Os, while the opposite situation 
occurs in $^{194,196}$Os. In $^{196}$Os the PES is already quite flat, 
evolving towards a vibrational shape in $^{200}$Os. The quadrupole 
moments for $^{188-200}$Os, shown in Fig. 14, also suggest a transition 
from prolate to oblate shapes between $^{192}$Os and $^{194}$Os.

\section{Discussion} 

The present calculations of Potential Energy Surfaces (PES) in the 
Relativistic Mean Field theory using the NL3 force lead to the following 
main conclusions.

a) The assumption made in the E(5) critical point symmetry that the potential 
in $\beta$ can be approximated by an infinite square well potential is 
justified, since rather flat PES are found for nuclei suggested as good 
examples of E(5). 

b) The PES found for nuclei suggested as good 
examples of X(5) exhibit a bump, in agreement with earlier calculations
\cite{Meng,ZhangSm}. The presence of a bump might be related to 
the success of the confined $\beta$-soft (CBS) rotor model
\cite{Pietr1,Pietr2}, the relevance of Davidson potentials in the description 
of X(5) propetries \cite{varPLB,varPRC}, as well as to the existence of 
a significant five-dimensional centrifugal effect found recently 
\cite{Caprio72}.  

c) The PES obtained for the series of Pt isotopes suggest a prolate to 
oblate shape transition at $^{186}$Pt.  

It is certainly of interest to calculate PES as functions of $\beta_2$ and 
$\gamma$, in order to examine to which extent the above conclusions are 
influenced by the $\gamma$ degree of freedom. In particular, one should 
examine carefully the PES in which two minima appear, since one of the minima 
could be a saddle point. 
In addition it is of interest to examine if and to which extent the present 
results will be changed if methods beyond the mean field approximation 
(see, for example, Refs. \cite{BBD.04,RER.04,ER.04}) are employed. 

\section*{\bf Acknowledgements} 

Partial support from the programme Pythagoras II of the Greek MNE \& RA under 
project 80861 (RF,GAL) and from the Greek State Scholarships Foundation 
(RF) is gratefully acknowledged.


\begin{figure}[ht]
\center{\includegraphics[height=160mm]{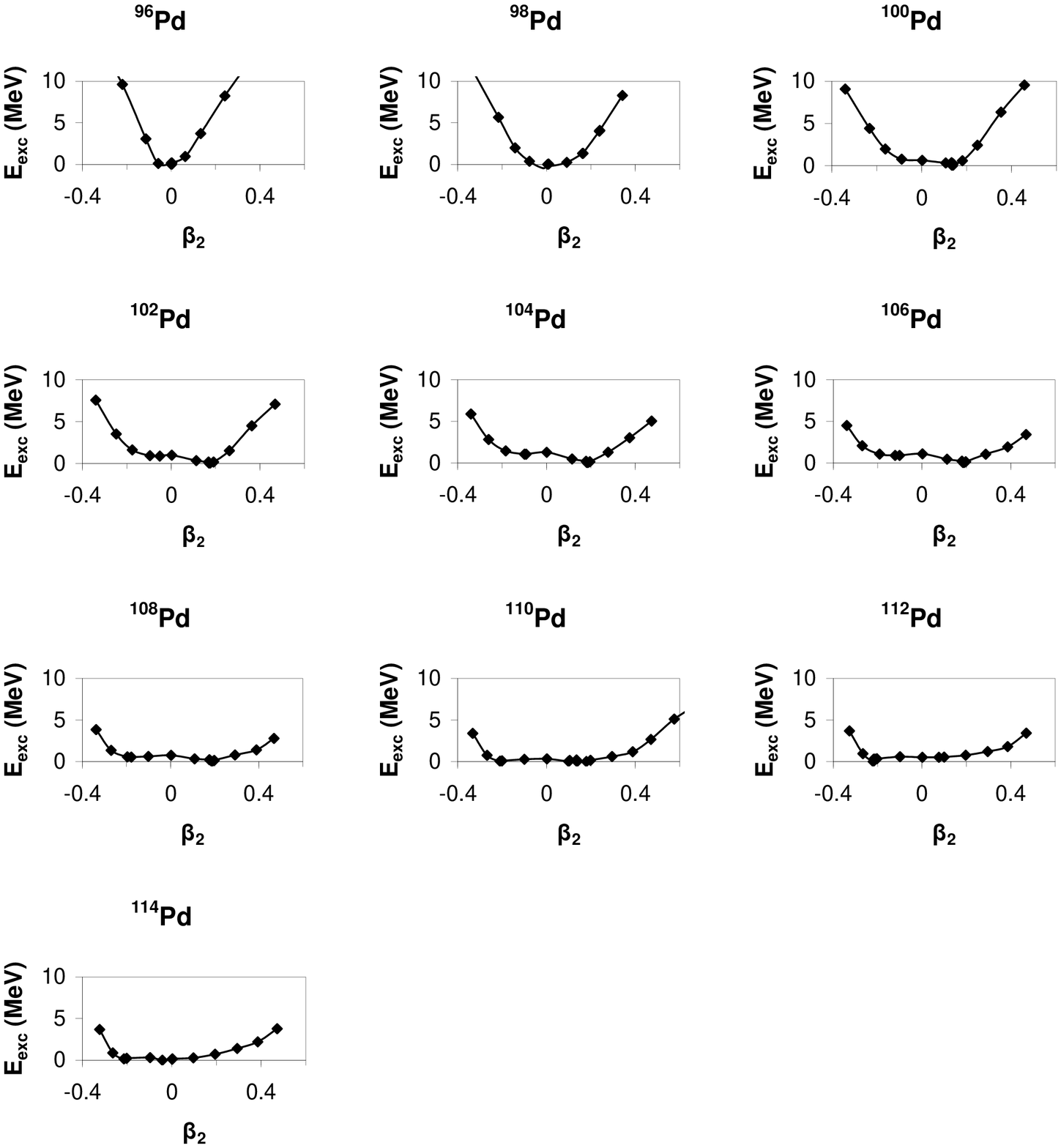}} 
\caption{(Color online) Potential energy surfaces (PES) for 
$^{96-114}$Pd, calculated using the Relativistic Hartree Bogoliubov theory 
with the NL3 force. }
\end{figure}  


\begin{figure}[ht]
\center{\includegraphics[height=120mm]{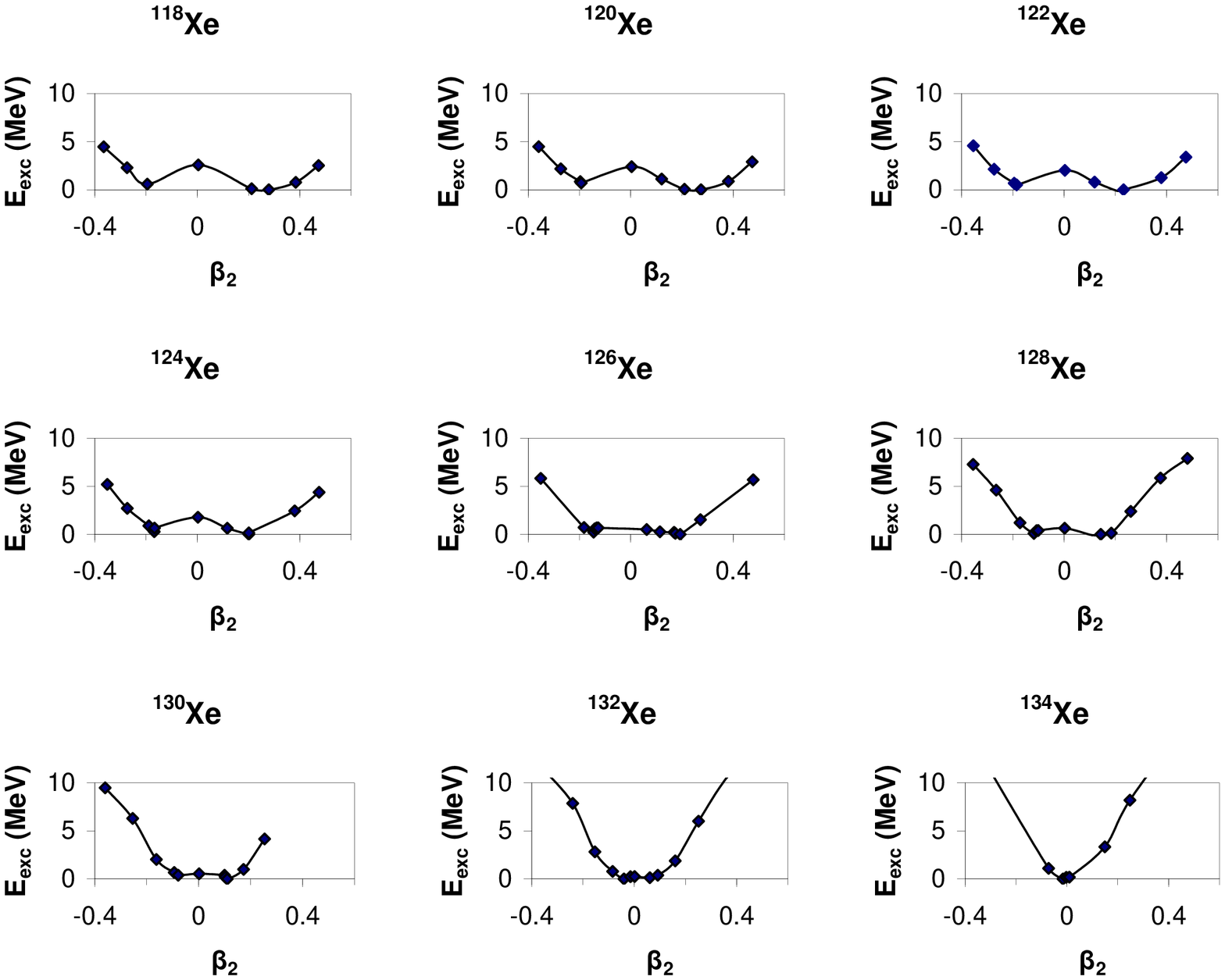}} 
\caption{(Color online) Same as Fig. 1, but for $^{118-134}$Xe.}
\end{figure}  


\begin{figure}[ht]
\center{\includegraphics[height=160mm]{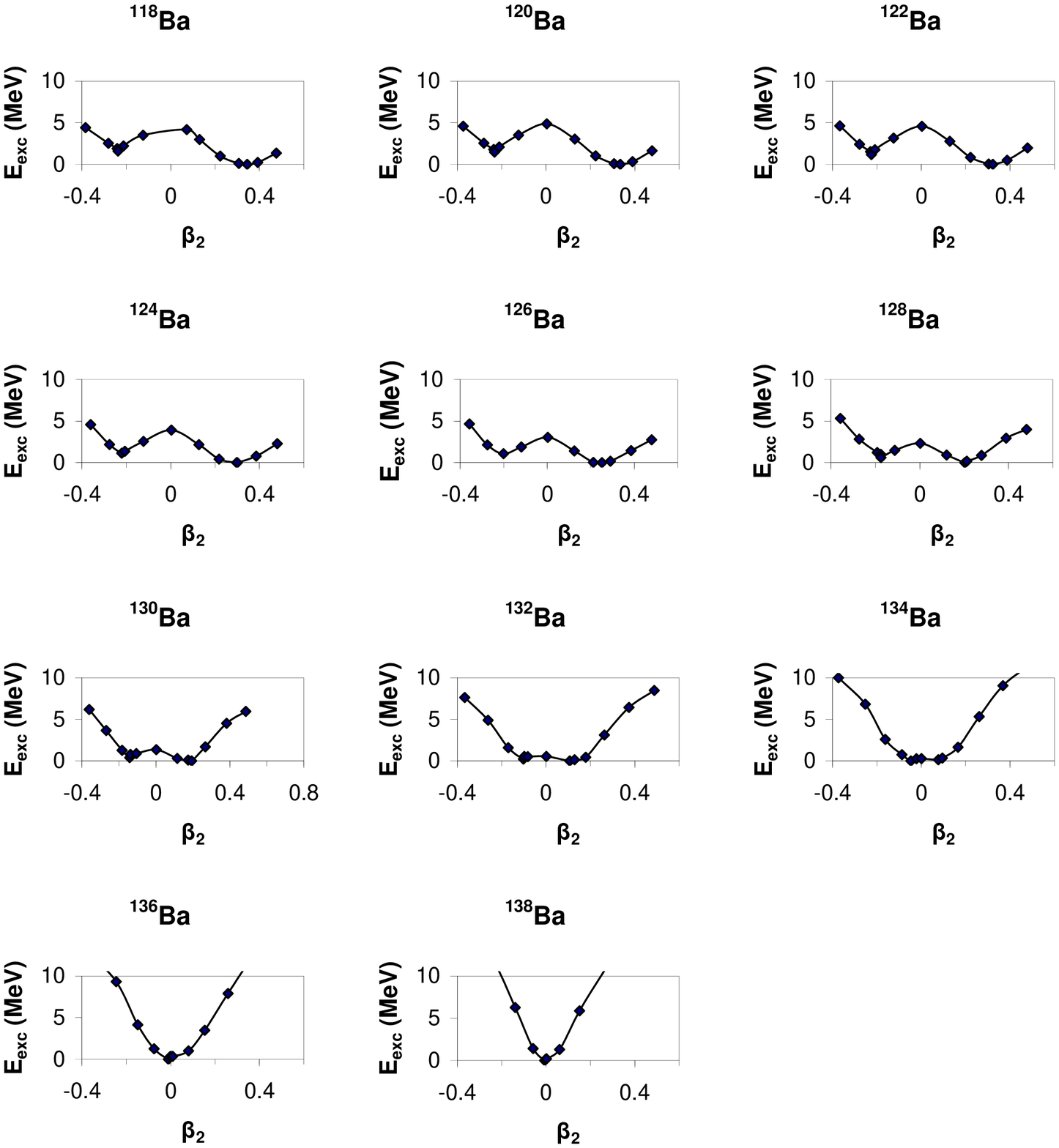}} 
\caption{(Color online) Same as Fig. 1, but for $^{118-138}$Ba.}
\end{figure}  


\begin{figure}[ht]
\center{\includegraphics[height=120mm]{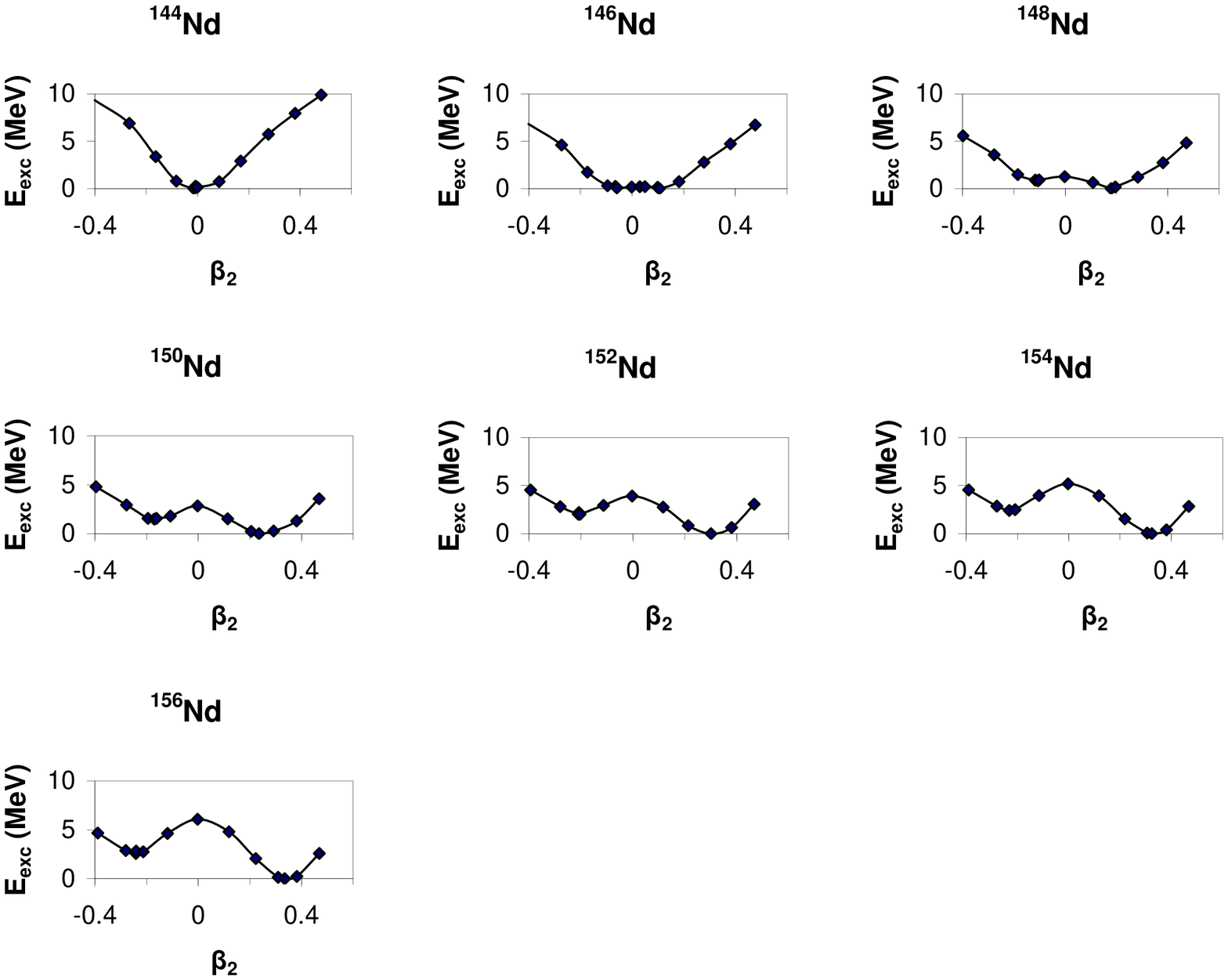}} 
\caption{(Color online) Same as Fig. 1, but for $^{144-156}$Nd.}
\end{figure}  


\begin{figure}[ht]
\center{\includegraphics[height=120mm]{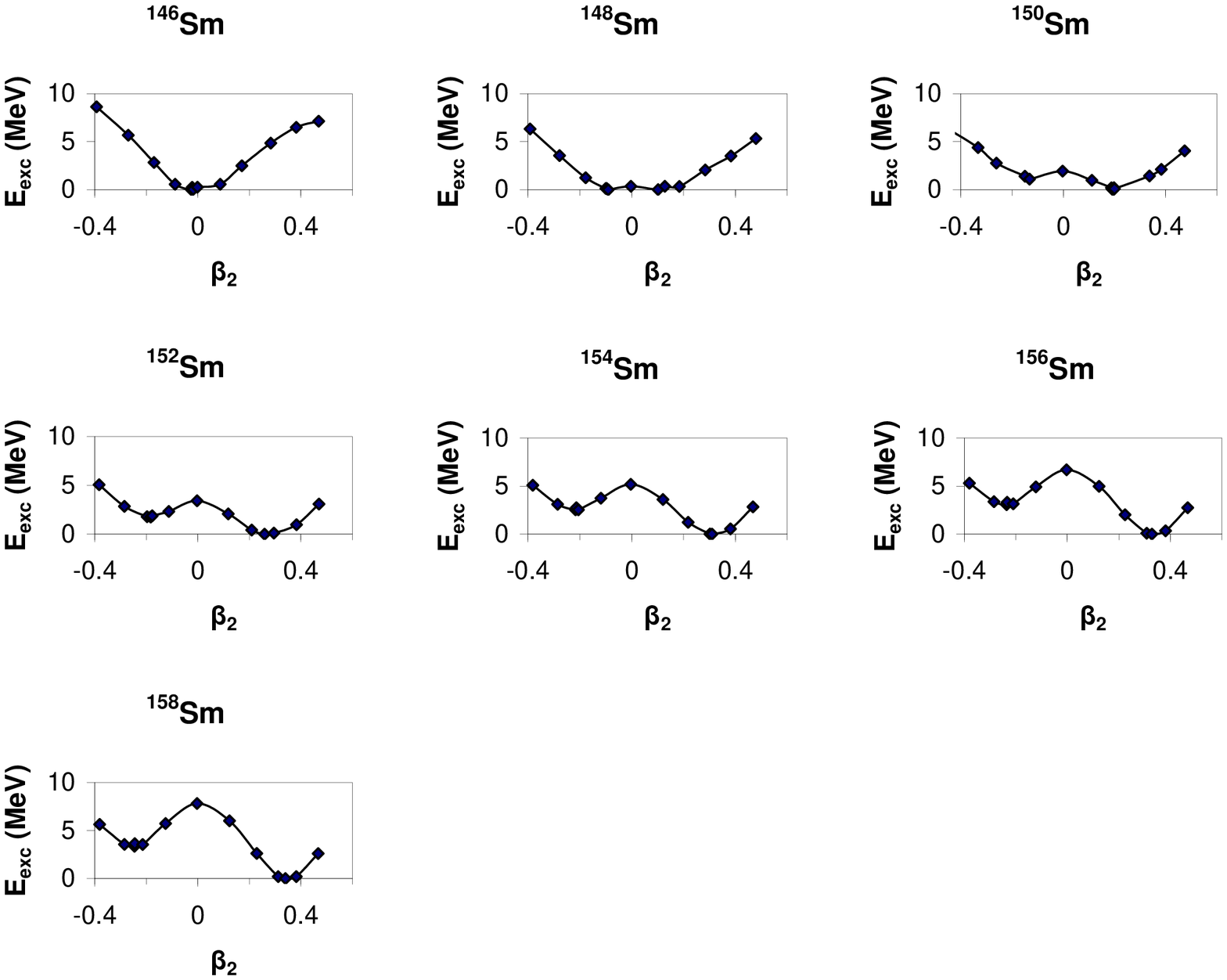}} 
\caption{(Color online) Same as Fig. 1, but for $^{146-158}$Sm.}
\end{figure}  


\begin{figure}[ht]
\center{\includegraphics[height=80mm]{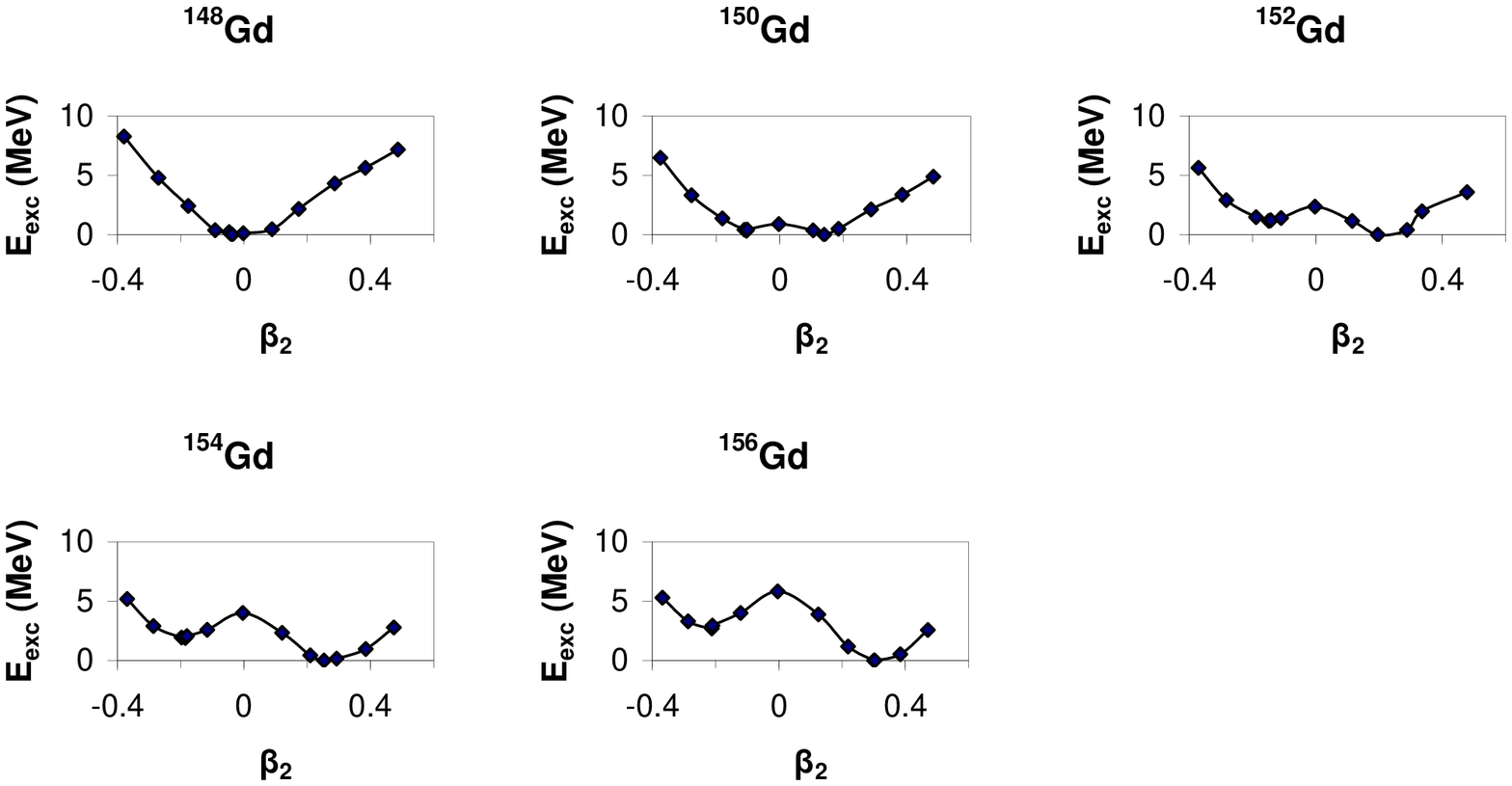}} 
\caption{(Color online) Same as Fig. 1, but for $^{148-156}$Gd.}
\end{figure}  


\begin{figure}[ht]
\center{\includegraphics[height=80mm]{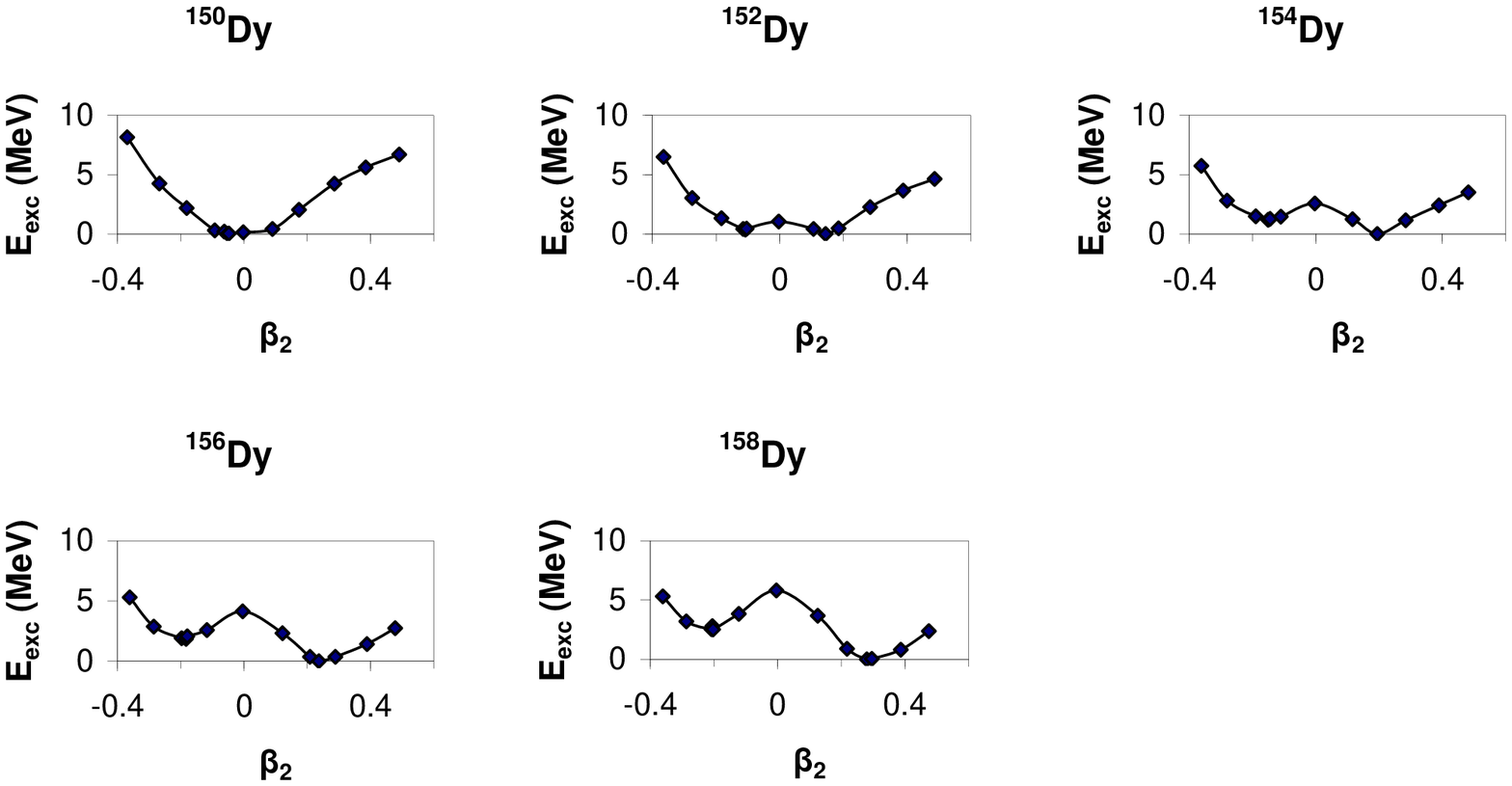}} 
\caption{(Color online) Same as Fig. 1, but for $^{150-158}$Dy.}
\end{figure}  


\begin{figure}[ht]
\center{\includegraphics[height=160mm]{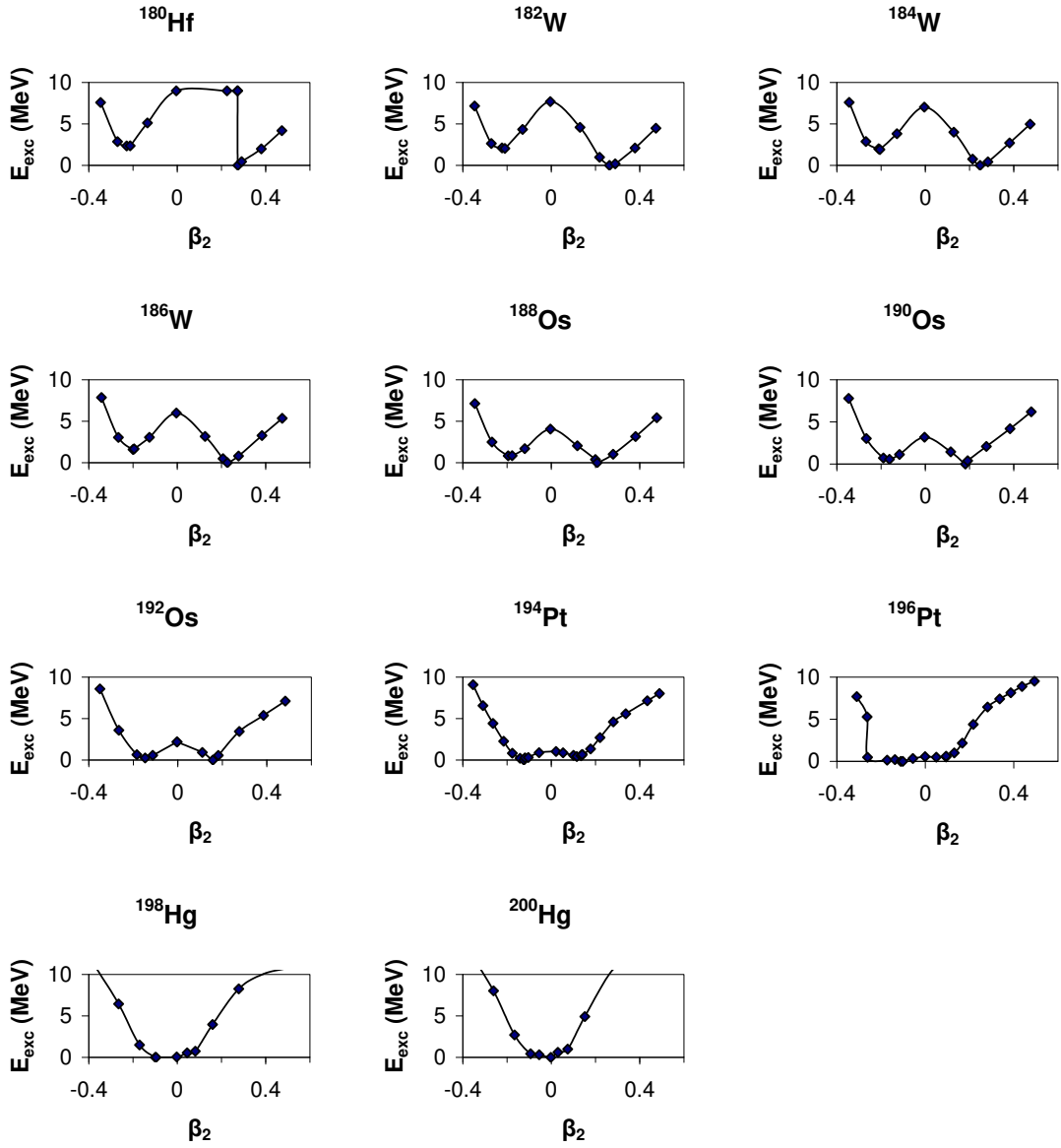}} 
\caption{(Color online) Same as Fig. 1, but for $^{180}$Hf, 
$^{182,184,186}$W, $^{188,190,192}$Os, $^{194,196}$Pt, $^{198,200}$Hg.}
\end{figure}  


\begin{figure}[ht]
\center{\includegraphics[height=50mm]{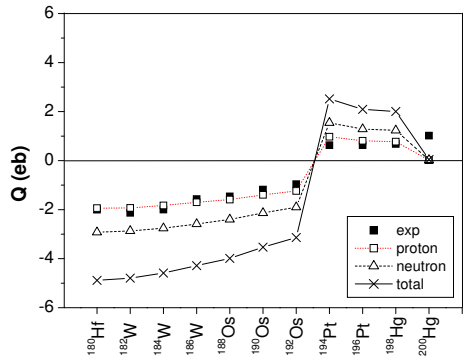}} 
\caption{ Quadrupole moments for the chain of nuclei of Fig. 8,
calculated using the Relativistic Hartree Bogoliubov theory with the NL3 
force. Experimental data are taken from Ref. \cite{NDS}.}
\end{figure}  


\begin{figure}[ht]
\center{\includegraphics[height=160mm]{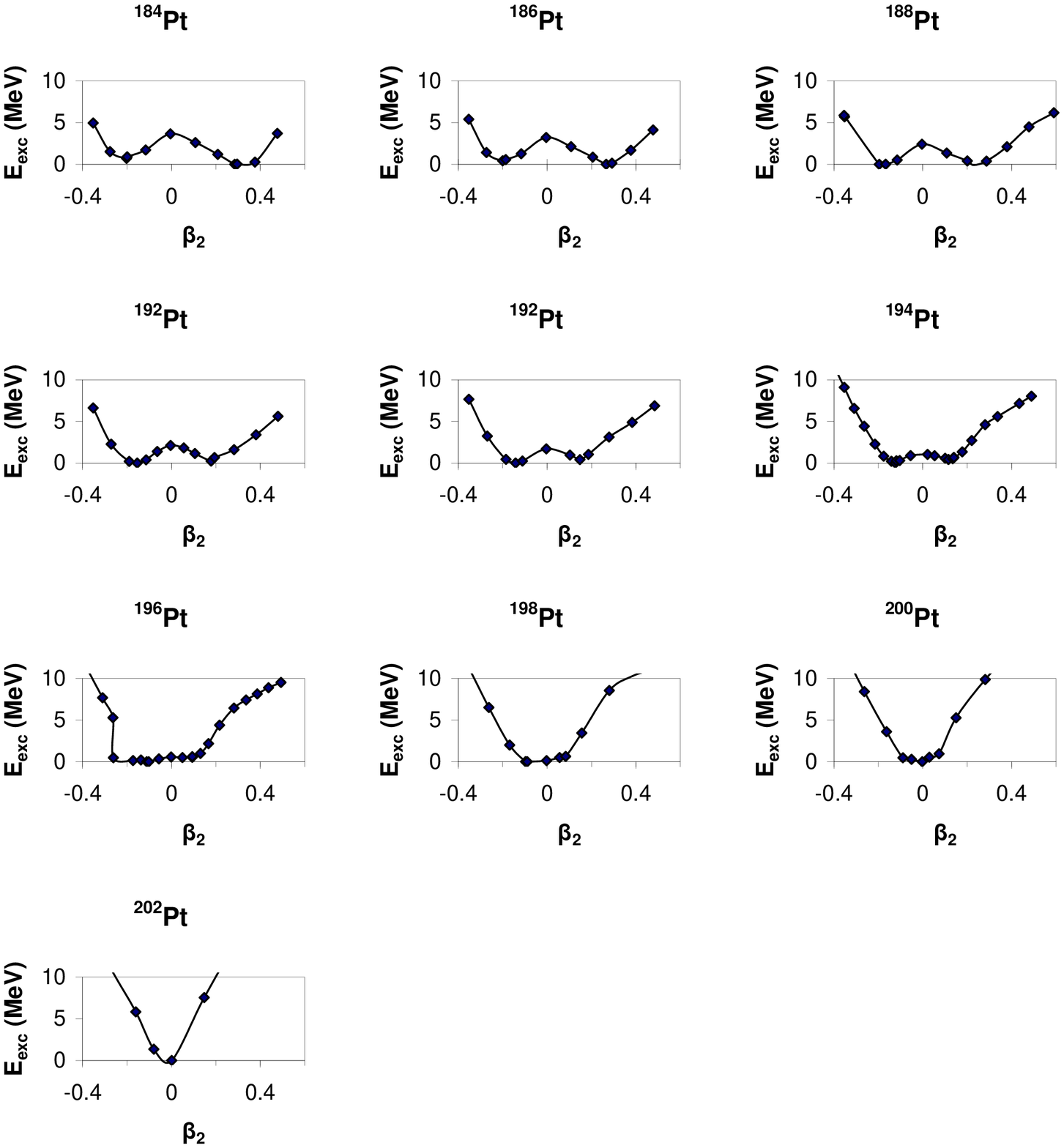}} 
\caption{(Color online) Same as Fig. 1, but for $^{184-202}$Pt.}
\end{figure}  


\begin{figure}[ht]
\center{\includegraphics[height=50mm]{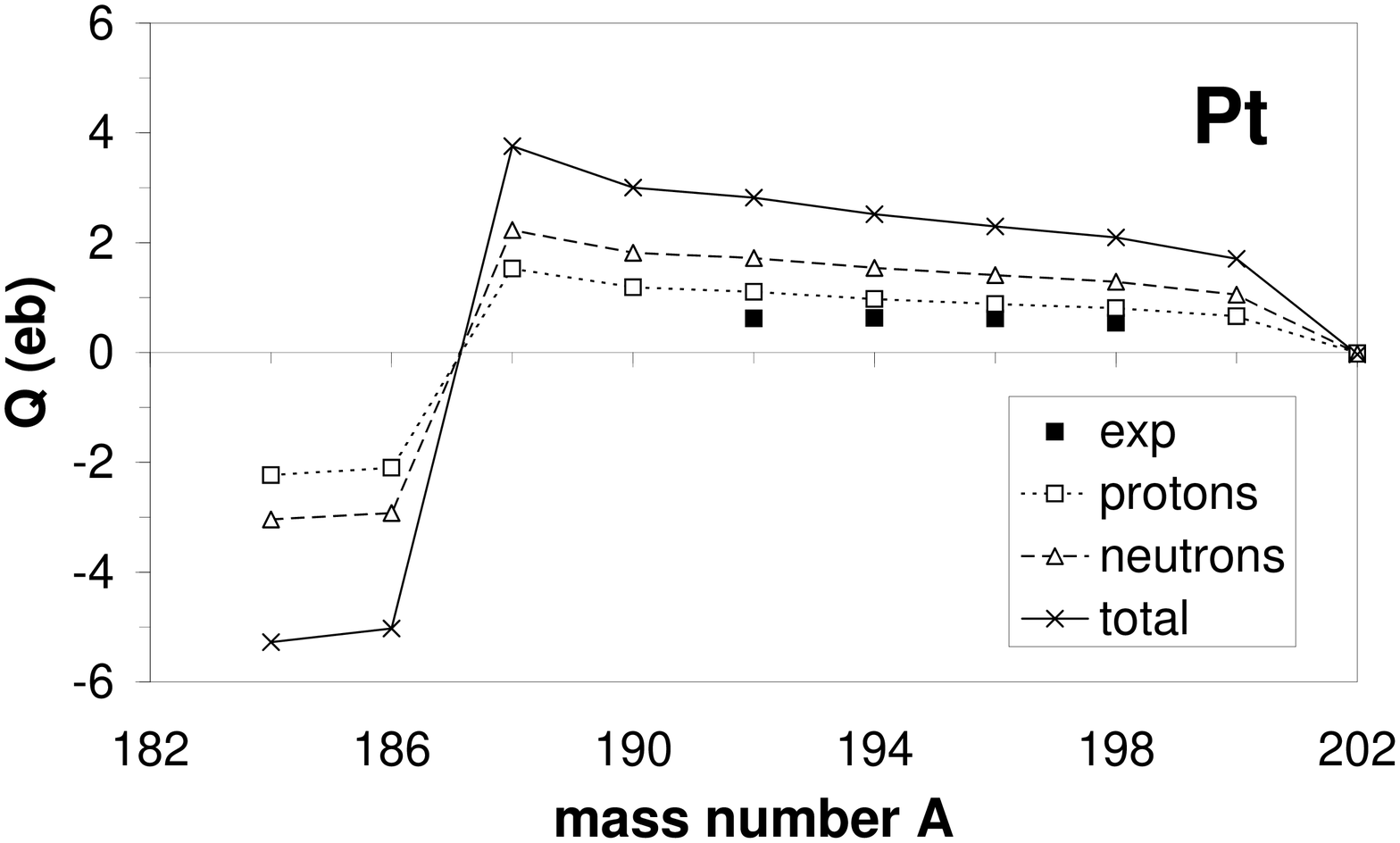}} 
\caption{ Same as Fig. 9, but for $^{184-202}$Pt.
Experimental data are taken from Ref. \cite{NDS}.}
\end{figure}  


\begin{figure}[ht]
\center{\rotatebox{270}{\includegraphics[height=80mm]{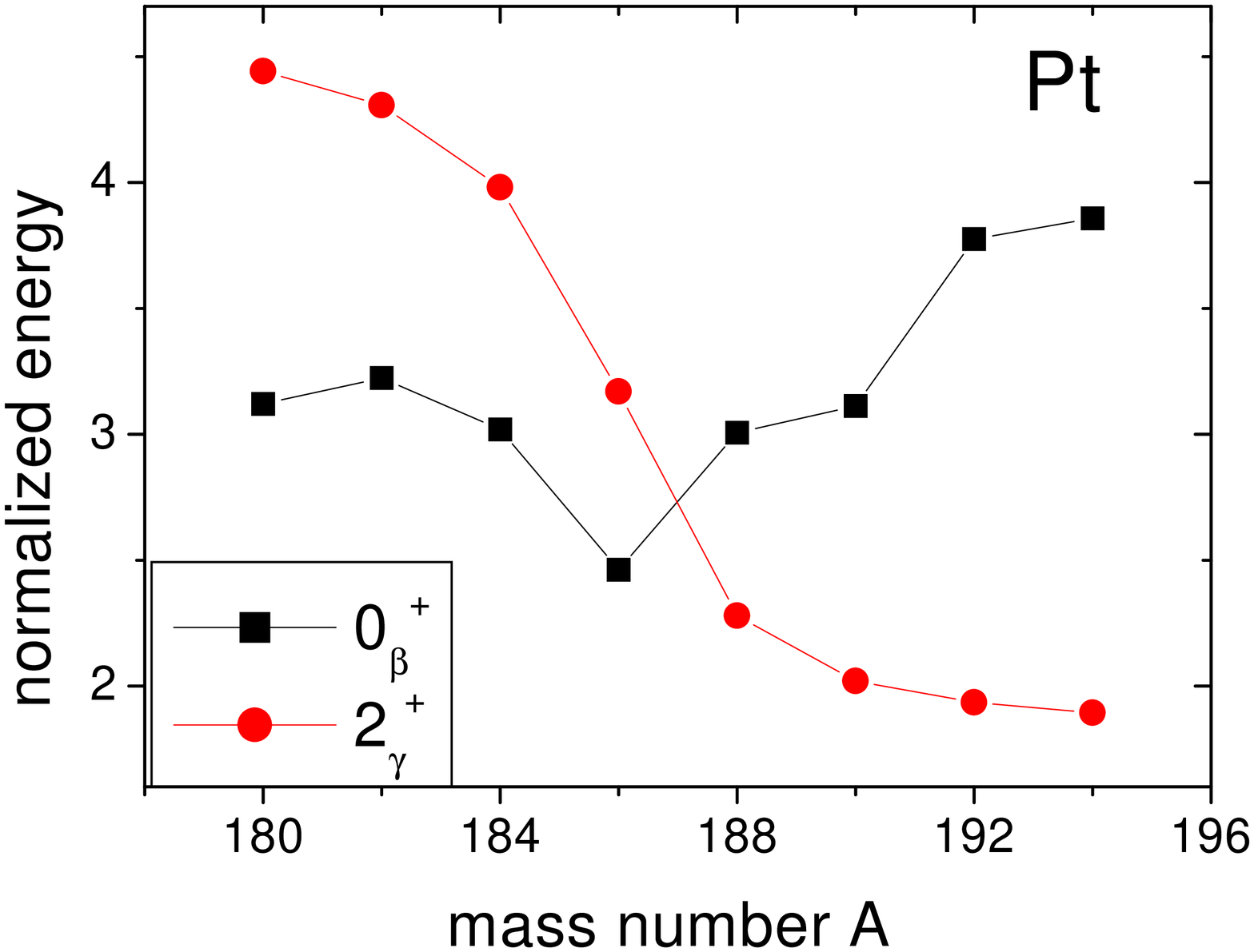}}} 
\caption{(Color online) Experimental energies of bandheads of the 
$\beta_1$ and $\gamma_1$ bands, normalized to the energy of the $2_1^+$ state 
of the ground state band, for several Pt isotopes. Experimental data are taken
from Ref. \cite{NDS}.}
\end{figure}  


\begin{figure}[ht]
\center{\includegraphics[height=120mm]{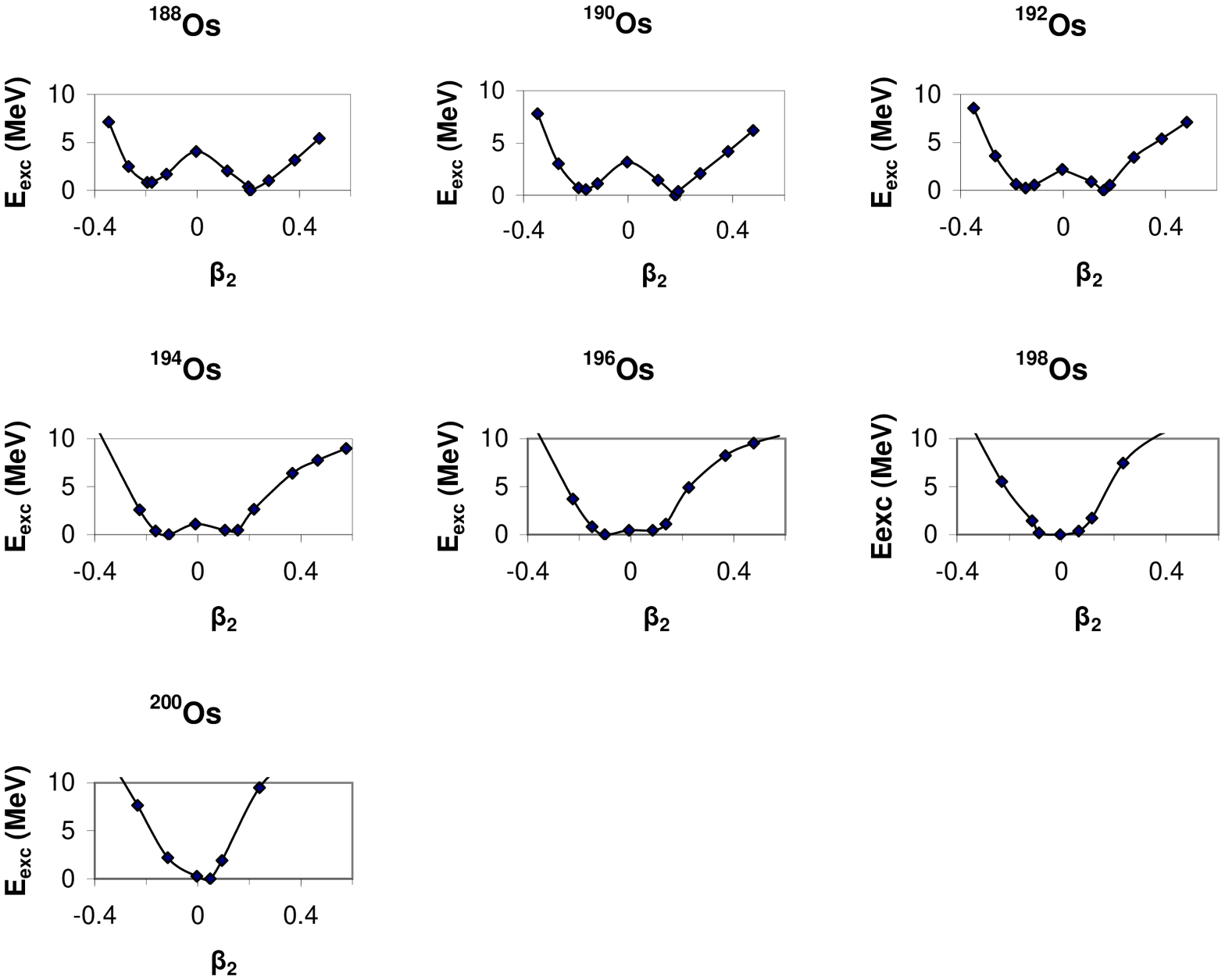}} 
\caption{(Color online) Same as Fig. 1, but for $^{188-200}$Os.}
\end{figure}  


\begin{figure}[ht]
\center{\includegraphics[height=50mm]{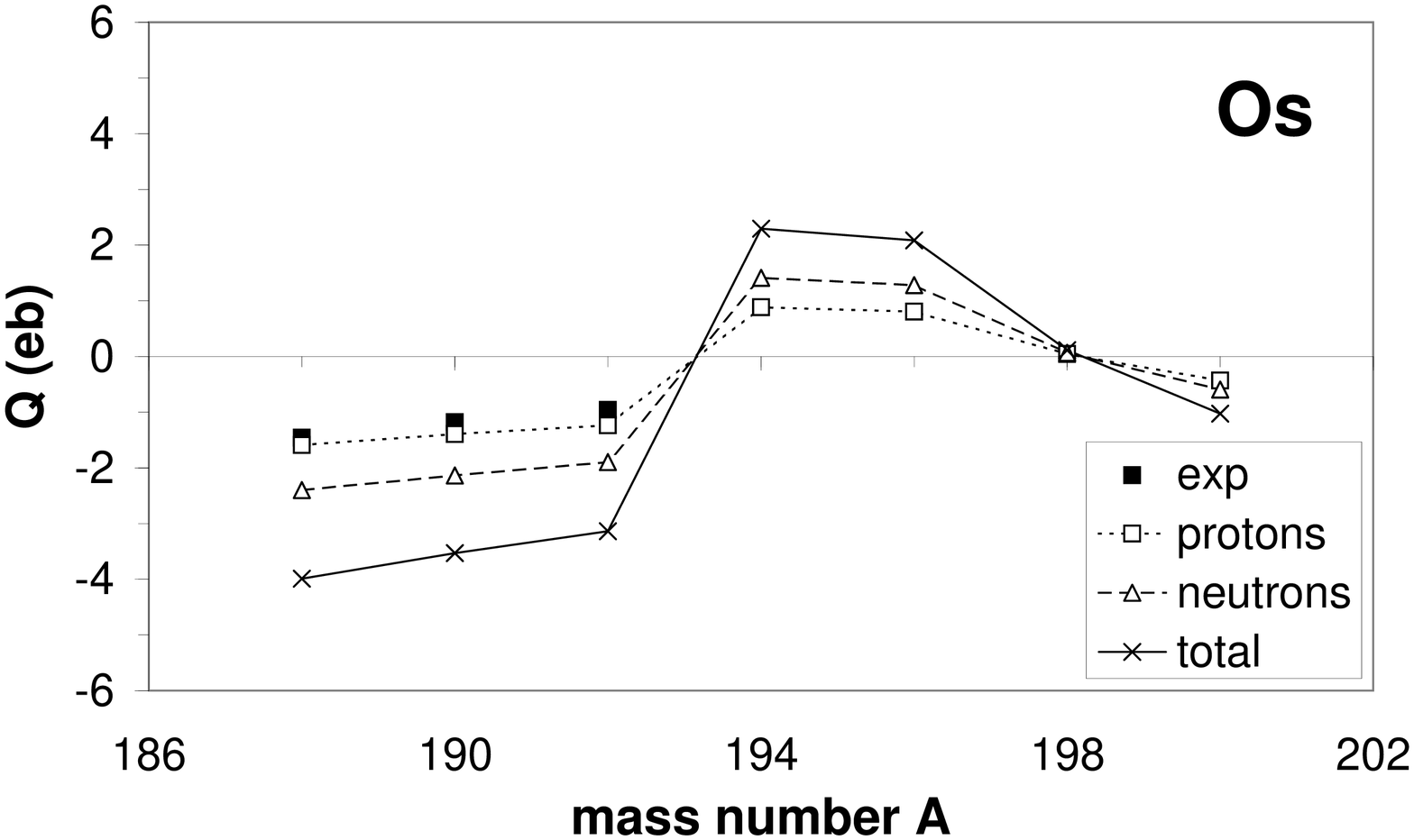}} 
\caption{ Same as Fig. 9, but for $^{188-200}$Os. 
Experimental data are taken from Ref. \cite{NDS}. }
\end{figure}  


\newpage 

\begin{thebibliography}{99}

\bibitem{IacE5}
F. Iachello, Phys. Rev. Lett. {\bf 85}, 3580 (2000). 

\bibitem{IacX5}
F. Iachello, Phys. Rev. Lett. {\bf 87}, 052502 (2001). 

\bibitem{CZE5}
R. F. Casten and N. V. Zamfir, Phys. Rev. Lett. {\bf 85}, 3584 (2000). 

\bibitem{CZX5} 
R. F. Casten and N. V. Zamfir, Phys. Rev. Lett. {\bf 87}, 052503 (2001).
 
\bibitem{ClarkE5}
R. M. Clark, M. Cromaz, M. A. Deleplanque, M. Descovich, R. M. Diamond, 
P. Fallon, I. Y. Lee, A. O. Macchiavelli, H. Mahmud, E. Rodriguez-Vieitez, 
F. S. Stephens, and D. Ward,  Phys. Rev. C {\bf 69}, 064322 (2004). 

\bibitem{ClarkX5}
R. M. Clark, M. Cromaz, M. A. Deleplanque, M. Descovich, R. M. Diamond, 
P. Fallon, R. B. Fierstone, I. Y. Lee, A. O. Macchiavelli, H. Mahmud, 
E. Rodriguez-Vieitez, F. S. Stephens, and D. Ward,  Phys. Rev. C {\bf 68},
037301 (2003).  

\bibitem{Bohr}
A. Bohr, Mat. Fys. Medd. K. Dan. Vidensk. Selsk.  {\bf 26}, no. 14 (1952).

\bibitem{Wilets}
L. Wilets and M. Jean, Phys. Rev. {\bf 102}, 788 (1956).

\bibitem{RoweI}
D. J. Rowe, Nucl. Phys. A {\bf 745}, 47 (2004). 

\bibitem{RoweII}
P. S. Turner and D. J. Rowe, Nucl. Phys. A {\bf 756}, 333 (2005). 

\bibitem{RoweIII}
G. Rosensteel and D. J. Rowe, Nucl. Phys. A {\bf 759}, 92 (2005). 

\bibitem{VALR.05}
D. Vretenar, A.~V. Afanasjev, G.~A. Lalazissis, and P. Ring, Physics
Reports {\bf 409}, 101 (2005).

\bibitem{LKR.97}
G.~A. Lalazissis, J. K{\"{o}}nig, and P. Ring, Phys. Rev. C {\bf 55}, 540
(1997).

\bibitem{Linnemann}
J. Jolie and A. Linnemann, Phys. Rev. C {\bf 68}, 031301 (2003).

\bibitem{Kucha.91} 
H. Kucharek and P. Ring, Z. Phys. A {\bf 339}, 23 (1991).

\bibitem{LA.99} 
G.~A. Lalazissis, D. Vretenar, and P. Ring, Nucl. Phys. A {\bf 650}, 133
(1999).

\bibitem{BGG.84} 
J.F. Berger, M. Girod and D. Gogny, Nucl. Phys. A {\bf 428}, 32 (1984).

\bibitem{Zamfir}
N. V. Zamfir, M. A. Caprio, R. F. Casten, C. J. Barton, C. W. Beausang, 
Z. Berant, D. S. Brenner, W. T. Chou, J. R. Cooper, A. A. Hecht, R. 
Kr\"ucken, H. Newman, J. R. Novak, N. Pietralla, A. Wolf, and K. E. Zyromski, 
Phys. Rev. C {\bf 65}, 044325 (2002).

\bibitem{Kalyva}
G. Kalyva, A. Spyrou, M. Axiotis, S. Harissopulos, A. Dewald, A. Fitzler, 
B. Saha, A. Linnemann, O. M\"oller, R. Vlastou, D. R. Napoli, N. Marginean, 
C. Rusu, G. de Angelis, C. Ur, D. Bazzacco, E. Farnea, D. L. Balabanski, 
and R. Julin, in {\it Frontiers in Nuclear Structure, Astrophysics and 
Reactions (Kos 2005)}, edited by R. Julin and S. Harissopulos (2005).

\bibitem{AriasE2}
J. M. Arias, Phys. Rev. C {\bf 63}, 034308 (2001). 

\bibitem{Zhang}
D.-L. Zhang and Y.-X. Liu, Phys. Rev. C {\bf 65}, 057301 (2002). 

\bibitem{E5}
D. Bonatsos, D. Lenis, N. Minkov, P. P. Raychev, and P. A. Terziev, 
Phys. Rev. C {\bf 69}, 044316 (2004). 

\bibitem{Ariasb4}
J. M. Arias, C. E. Alonso, A. Vitturi, J. E. Garc\'{\i}a-Ramos, J. Dukelsky, 
and A. Frank, Phys. Rev. C {\bf 68}, 041302 (2003). 

\bibitem{Ariasb4b}
J. E. Garc\'{\i}a-Ramos, J. Dukelsky, and J. M. Arias, Phys. Rev. C {\bf 72}, 
037301 (2005).  

\bibitem{Kirson}
M. W. Kirson, Phys. Rev. C {\bf 70}, 049801 (2004). 

\bibitem{Liu}
D.-L. Zhang and Y.-X. Liu, Chin. Phys. Lett. {\bf 20}, 1028 (2003). 

\bibitem{Kneissl}
U. Kneissl, in {\it Key Topics in Nuclear Structure (Paestum 2004)}, edited by
A. Covello (World Scientific, Singapore, 2005) p. 399.  

\bibitem{NDS}
Nuclear Data Sheets, as of June 2005. 

\bibitem{IBM}
F. Iachello and A. Arima, {\it The Interacting Boson Model} (Cambridge 
University Press, Cambridge, 1987). 

\bibitem{Kruecken} 
R. Kr\"ucken, B. Albanna, C. Bialik, R. F. Casten, J. R. Cooper, A. Dewald, 
N. V. Zamfir, C. J. Barton, C. W. Beausang, M. A. Caprio, A. A. Hecht, 
T. Klug, J. R. Novak, N. Pietralla, and P. von Brentano, Phys. Rev. Lett. 
{\bf 88}, 232501 (2002).

\bibitem{ZamfirSm}
N. V. Zamfir, H. G. B\"orner, N. Pietralla, R. F. Casten, Z. Berant, 
C. J. Barton, C. W. Beausang, D. S. Brenner, M. A. Caprio, J. R. Cooper, 
A. A. Hecht, K. Krti\v{c}ka, R. Kr\"ucken, P. Mutti, J. R. Novak, and A. Wolf, 
Phys. Rev. C {\bf 65}, 067305 (2002). 

\bibitem{Clark}
R. M. Clark, M. Cromaz, M. A. Deleplanque, R. M. Diamond, P. Fallon, 
A. G\"orgen, I. Y. Lee, A. O. Macchiavelli, F. S. Stephens, and D. Ward,   
Phys. Rev. C {\bf 67}, 041302 (2003). 

\bibitem{CZK}
R. F. Casten, N. V. Zamfir, and R. Kr\"ucken, Phys. Rev. C {\bf 68}, 059801 
(2003). 

\bibitem{Bijker}
R. Bijker, R. F. Casten, N. V. Zamfir, and E. A. McCutchan, Phys. Rev. C 
{\bf 68}, 064304 (2003). Erratum: Phys. Rev. C {\bf 69}, 059901 (2004). 

\bibitem{Zhao}
D.-L. Zhang and H.-Y. Zhao, Chin. Phys. Lett. {\bf 19}, 779 (2002). 

\bibitem{Tonev} 
D. Tonev, A. Dewald, T. Klug, P. Petkov, J. Jolie, A. Fitzler, O. M\"oller, 
S. Heinze, P. von Brentano, and R. F. Casten, Phys. Rev. C {\bf 69},
034334 (2004).

\bibitem{Dewald}
A. Dewald, O. M\"oller, D. Tonev, A. Fitzler, B. Saha, K. Jessen, S. Heinze, 
A. Linnemann, J. Jolie, K. O. Zell, P. von Brentano, P. Petkov, R. F. Casten,
M. Caprio, J. R. Cooper, R. Kr\"ucken, V. Zamfir, D. Bazzacco, S. Lunardi, 
C. Rossi Alvarez, F. Brandolini, C. Ur, G. de Angelis, D. R. Napoli, E. Farnea,
N. Marginean, T. Martinez, and M. Axiotis, Eur. Phys. J A {\bf 20}, 173 (2004).
 
\bibitem{CaprioDy}
M. A. Caprio, N. V. Zamfir, R. F. Casten, C. J. Barton, C. W. Beausang, 
J. R. Cooper, A. A. Hecht, R. Kr\"ucken, H. Newman, J. R. Novak, N. Pietralla,
A. Wolf, and K. E. Zyromski, Phys. Rev. C {\bf 66}, 054310 (2002). 

\bibitem{Fransen}
C. Fransen, N. Pietralla, A. Linnemann, V. Werner, and R. Bijker, 
Phys. Rev. C {\bf 69}, 014313 (2004). 

\bibitem{X5}
D. Bonatsos, D. Lenis, N. Minkov, P. P. Raychev, and P. A. Terziev, 
Phys. Rev. C {\bf 69}, 014302 (2004).

\bibitem{Meng}
J. Meng, W. Zhang, S. G. Zhou, H. Toki, and L. S. Geng, Eur. Phys. J. 
A {\bf 25}, 23 (2005).

\bibitem{ZhangSm}
J.-Y. Zhang, M. A. Caprio, N. V. Zamfir, and R. F. Casten, Phys. Rev. C 
{\bf 60}, 061304 (1999).

\bibitem{Pietr1}
N. Pietralla and O. M. Gorbachenko, Phys. Rev. C {\bf 70}, 011304 (2004). 

\bibitem{Pietr2}
K. Dusling and N. Pietralla, Phys. Rev. C {\bf 72}, 011303 (2005). 

\bibitem{Dav}
P. M. Davidson, Proc. R. Soc. {\bf 135}, 459 (1932).  

\bibitem{Rowe}
D. J. Rowe and C. Bahri, J. Phys. A  {\bf 31}, 4947 (1998). 

\bibitem{varPLB}
D. Bonatsos, D. Lenis, N. Minkov, D. Petrellis, P. P. Raychev, and P. A. 
Terziev, Phys. Lett. B {\bf 584}, 40 (2004). 

\bibitem{varPRC}
D. Bonatsos, D. Lenis, N. Minkov, D. Petrellis, P. P. Raychev, and P. A. 
Terziev, Phys. Rev. C {\bf 70}, 024305 (2004). 

\bibitem{Caprio72}
M. A. Caprio, Phys. Rev. C {\bf 72}, 054323 (2005). 

\bibitem{Rowe735}
D. J. Rowe, Nucl. Phys. A {\bf 735}, 372 (2004). 

\bibitem{Rowe45}
D. J. Rowe, P. S. Turner, and J. Repka, J. Math. Phys. {\bf 45}, 2761 (2004).

\bibitem{Rowe753}
D. J. Rowe and P. S. Turner, Nucl. Phys. A {\bf 753}, 94 (2005). 

\bibitem{Casten}
R. F. Casten, {\it Nuclear Structure from a Simple Perspective}
(Oxford University Press, Oxford, 1990).

\bibitem{McCPt} 
E. A. McCutchan, R. F. Casten, and N. V. Zamfir, Phys. Rev. C {\bf 71},
061301 (2005). 

\bibitem{IZC}
F. Iachello, N. V. Zamfir, and R. F. Casten, Phys. Rev. Lett. {\bf 81},
1191 (1998). 

\bibitem{McCZC}
E. A. McCutchan, N. V. Zamfir, and R. F. Casten, Phys. Rev. C {\bf 69}, 
064306 (2004). 

\bibitem{BBD.04}
M. Bender, P. Bonche, T. Duguet, and P.-H. Heenen, Phys. Rev. C {\bf 69},
064303  (2004).

\bibitem{RER.04}
R.~R. Rodriguez-Guzman, J.~L. Egido, and L.~M. Robledo, Phys. Rev. C {\bf 69},
054319  (2004).

\bibitem{ER.04}
J. L. Egido and L. M. Robledo,  in {\it Lecture Notes in Physics},
edited by G. Lalazissis, P. Ring, and D. Vretenar (Springer-Verlag, 
Heidelberg, 2004), Vol. {\bf 641}, p. 269.

\end{thebibliography}
\end{document}